\documentclass[prl, twocolumn]{revtex4-1}
\usepackage{bibunits}
\usepackage{amsmath,amssymb,bm}
\usepackage{graphicx}
\usepackage{epstopdf}
\usepackage{latexsym}
\usepackage{subfigure}
\usepackage{color}
\usepackage{natbib}
\usepackage{braket}
\usepackage{hyperref}
\hypersetup{
  colorlinks,
  citecolor=blue,
  linkcolor=blue,
  urlcolor=blue}
      
\begin{document}

\title{Spin nematics, valence-bond solids and spin liquids in 
  SO($N$) quantum spin models on the triangular lattice}
\author{Ribhu K. Kaul}
\affiliation{Department of Physics \& Astronomy, University of Kentucky, Lexington, KY-40506-0055}
\begin{abstract}
We introduce a simple model of SO($N$) spins with two-site
interactions which is amenable to quantum Monte-Carlo studies without a sign problem
on {\em non-bipartite} lattices.  We present numerical results for
this model on the two-dimensional
triangular lattice where we find evidence for a spin nematic at
small $N$, 
a valence-bond solid (VBS) at large $N$ and a quantum spin
liquid at intermediate $N$. By the introduction of a sign-free
four-site interaction we 
uncover a rich phase diagram with evidence for both first-order and exotic
 continuous phase transitions.

\end{abstract}
\maketitle

\begin{bibunit}[apsrev]

The destruction of magnetic order by quantum fluctuations in
spin systems
is frequently invoked
as a route to exotic condensed matter physics such as spin liquid phases and novel
quantum critical points~\cite{balents2010:spliq,sachdev1999:qpt,xu2012:qcp}. 
The most commonly studied spin Hamiltonians have symmetries of the groups
SO(3) and SU(2) which describe the rotational symmetry of 3-dimensional
space. Motivated both by theoretical and 
experimental~\cite{Wu2010:ExoticManyBody} interest, spin models with larger-$N$
symmetries have been introduced, e.g. extensions of SU(2) to
SU($N$)~\cite{sutherland1975:sun,affleck1985:lgN,read1989:vbs,kaul2013:qmc} or
Sp($N$)~\cite{read1991:spN}. 

The extension of SO(3) to SO($N$) is an independant
large-$N$ enlargement of symmetry, with its own physical motivations~\cite{demler2004:so5}. 
While there have been many studies of SO($N$)
spin models in one
dimension~\cite{tu2008:son,alet2011:so5,okunishi2014:son}, our
understanding of their ground states and quantum phase
transitions in higher dimension is in its infancy.
To this end, we introduce here a simple  SO($N$) spin
model that surprisingly is sign free on
{\em any} non-bipartite lattice. This model provides us with a new setting in
which the destruction of magnetic order can be studied in higher
dimensions using {\em unbiased} methods. As an example of interest,
we present the results of a detailed study of the phase diagram of the our
SO($N$) anti-ferromagnet on the two-dimensional triangular lattice.

{\em Models. --} 
Consider a triangular lattice, each site of which has a Hilbert state of $N$ states, we
will denote the state of site $j$ as $|\alpha\rangle_j$ ($1\leq
\alpha\leq N$). Define the $N(N-1)/2$ generators of SO($N$) on site
$i$ as  $\hat L^{\alpha\beta}_i$ with $\alpha<\beta$; they will be chosen in
the fundamental representation on all sites:
$\hat L^{\alpha\beta}_j | \gamma \rangle_j = i \delta_{\beta \gamma}|\alpha
\rangle_j - i \delta_{\alpha\gamma}|\beta\rangle_j$. Now consider the
following SO($N$)~\cite{nbdrct2015:symm} symmetric lattice model for $N\geq 3$,
\begin{equation}
\label{eq:jmodel}
 \hat H_J = - \frac{J}{N^2 -2N} \sum_{\langle ij \rangle}(\hat L_i\cdot \hat L_j)^2,
\end{equation}
where the ``$\cdot$'' implies a summation over the $N(N-1)/2$ generators
and $\langle ij \rangle$ is the set of nearest neighbors. To see that $\hat H_J$ does
not suffer from the sign problem, define a ``singlet'' state on a bond,  $|S_{ij}\rangle \equiv
\frac{1}{\sqrt{N}}\sum_\alpha|\alpha \alpha \rangle_{ij}$ and the
singlet projector $\hat P_{ij}=|S_{ij}\rangle \langle S_{ij}|$. Using these
operators and ignoring a constant shift we find the simple form~\cite{nbdrct2015:supmat},
\begin{equation}
\label{eq:projmodel}
\hat H_J = - J\sum_{\langle ij \rangle} \hat P_{ij}.
\end{equation}
We make four observations: First, it is possible to create an SO($N$) spin
singlet with only two spins for all $N$ (in contrast to SU($N$) where
$N$ fundamental spins are required to create a singlet); Second Eq.~(\ref{eq:jmodel}) being
a sum of projectors on this two-site singlet is the simplest SO($N$)
coupling, despite it being a biquadratic interaction in the generators
$\hat L^{\alpha\beta}$;
Third, since the singlet has a positive expansion, $\hat H_J$ is Marshall
positive on {\em any} lattice;
Fourth, on bipartite lattices $\hat H_J$ is equivalent to the familiar SU($N$) 
anti-ferromagnet~\cite{affleck1985:lgN}, i.e. the obvious SO($N$) of Eq.~(\ref{eq:jmodel}) is enlarged to
an SU($N$) symmetry. Since the bipartite SU($N$) case has been
studied in great detail in past work on various lattices~\cite{read1989:vbs,santoro1999:sun,harada2003:sun,beach2009:sun,lou2009:sun,
  kaul2012:bilayer,kaul2012:j1j2,block2012:cubic,block2013:fate}, we shall concern ourselves here with the
non-bipartite SO($N$) case which is relatively unexplored.

{\em Phases of ${\hat H}_J$:} 
Starting at $N=3$,  Eq.~(\ref{eq:jmodel}) becomes
 $\hat H= -\frac{J}{3} \sum_{\langle ij\rangle}\left ( \vec S_i\cdot \vec S_j \right
 )^2$ with $\vec{S}$ the familiar $S=1$ representation of angular
 momentum. Previous numerical work has shown that this triangular
 lattice $S=1$ biquadratic
 model~\cite{kaul2012:biq, laeuchli2006:nigas} has an SO(3)
symmetry breaking ``spin nematic'' magnetic ground state (we shall
denote this phase by SN). The ground state of $\hat H_J$ for
$N> 3$ has not been studied in the past. 

In
the large-$N$ limit, analogous to previous work for SU($N$)
anti-ferromagnets on bipartite lattices~\cite{read1989:nucphysB},
 the ground state is
infinitely degenerate and consists of dimer coverings where each dimer
is in $|S_{ij}\rangle$. At leading order in
$1/N$, $\hat H_J$ introduces off-diagonal moves which re-arrange parallel dimers around
a plaquette, mapping $\hat H_J$ at large-$N$ to a quantum dimer model on the triangular
lattice with only a kinetic term, 
\begin{equation}
\hat{H}_{\rm QDM} = -t
\sum_{\rm plaq}\left\{ 
\left(|\setlength{\unitlength}{3158sp}%
\begingroup\makeatletter\ifx\SetFigFont\undefined%
\gdef\SetFigFont#1#2#3#4#5{%
  \reset@font\fontsize{#1}{#2pt}%
  \fontfamily{#3}\fontseries{#4}\fontshape{#5}%
  \selectfont}%
\fi\endgroup%
\begin{picture}(319,210)(517,-186)
\thinlines
\thicklines
\put(720, -9){\circle{28}}
\put(803,-153){\line(-1, 0){171}}
\put(547, -9){\circle{28}}
\put(722, -9){\line(-1, 0){172}}
\put(800,-152){\circle{28}}
\put(627,-152){\circle{28}}
\end{picture}
\rangle_i
\langle \setlength{\unitlength}{3158sp}%
\begingroup\makeatletter\ifx\SetFigFont\undefined%
\gdef\SetFigFont#1#2#3#4#5{%
  \reset@font\fontsize{#1}{#2pt}%
  \fontfamily{#3}\fontseries{#4}\fontshape{#5}%
  \selectfont}%
\fi\endgroup%
\begin{picture}(317,216)(525,-421)
\thinlines
\thicklines
\put(643,-386){\circle{28}}
\multiput(809,-386)(-5.48713,9.14522){17}{\makebox(8.3333,12.5000){\SetFigFont{7}{8.4}{\rmdefault}{\mddefault}{\updefault}.}}
\put(556,-237){\circle{28}}
\multiput(644,-388)(-5.53125,9.21875){17}{\makebox(8.3333,12.5000){\SetFigFont{7}{8.4}{\rmdefault}{\mddefault}{\updefault}.}}
\put(807,-385){\circle{28}}
\put(720,-234){\circle{28}}
\end{picture}
|_i+h.c. \right) \right. 
\label{eq:qdm}
\end{equation}
where the sum on plaquettes includes all closed loops of length four
on the triangular lattice.
The ground state of this model has
been found in previous analytic~\cite{moessner2001:isfr} and
numerical work~\cite{ralko2006:qdm} to be a
$\sqrt{12}\times \sqrt{12}$ valence bond solid (VBS), breaking the
lattice translation symmetry. We thus expect that
at large but finite values of $N$, $\hat H_J$ should restore its SO($N$) symmetry and enter this same VBS state.

\begin{figure}[!t]
\includegraphics[width=3.5in,trim=0 170 0 0,clip=true]{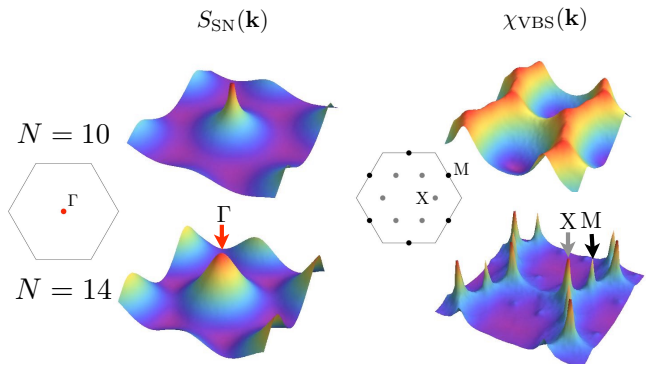}
\caption{\label{fig:Sk} Equal time structure factors for SN order [$S_{\rm
    SN}({\bf k})$], and susceptibility for VBS order [$\chi_{\rm VBS}({\bf k})$] shown for $N=10$
  and $N=14$, for the $\hat H_J$ model, Eq.~(\ref{eq:jmodel}) with $L=48$. The Bragg peaks for SN (VBS) weaken (sharpen) with
  increasing $N$. The cartoon of the Brillouin zones shows the
  location of the ordering
  vectors of both order parameters. Quantitative finite size scaling
  of these orders is shown in Fig.~\ref{fig:R_MV}.  }
\end{figure}

Since $\hat H_J$ has SN order for $N=3$ and is
expected to have a non-magnetic VBS at large-$N$, it is  interesting to ask what the nature
of the transition at which SN magnetism is destroyed. The answer to this
question is unclear based on current theoretical ideas and
is best settled by unbiased numerical simulations. Exploiting that $\hat H_J$ has no sign
problem we study it as a function of $N$ on $L\times L$
lattices at temperture $\beta$
by unbiased stochastic series expansion~\cite{sandvik2010:vietri}
quantum Monte Carlo simulations, with a previously described algorithm
~\cite{kaul2012:biq}. The SN state is described by the matrix order
parameter $\hat Q_{\alpha\beta} = |\alpha\rangle\langle\beta| - \frac{1}{N}$. The static structure factor, $S_{\rm SN} ({\bf
k}) = \frac{1}{N_{\rm site}}  \sum_{ij } e^{i {\bf k}\cdot ({\bf
  r}_i-{\bf r}_j)}\langle \hat Q_{\alpha\alpha}(i)
\hat Q_{\alpha\alpha}(j) \rangle$ is used to detect SN order. For the VBS
order, we construct the
${\bf k}$ dependent susceptibility of dimer-dimer
correlation functions in the
usual way from imaginary time-displaced operators:  $\chi_{\rm VBS} ({\bf
k}) = \frac{1}{N_{\rm site}}  \sum_{ij } e^{i {\bf k}\cdot ({\bf
  r}_i-{\bf r}_j)} \frac{1}{\beta}\int d \tau \langle \hat P_{{\bf
  r}_i,{\bf r}_i+{\hat x}}(\tau) \hat P_{{\bf r}_j,{\bf r}_j+{\hat
  x}}(0)\rangle $. Throughout this paper we have fixed $\beta=L$ for
our finite size scaling~\cite{nbdrct2015:supmat}.

\begin{figure}[!t]
\includegraphics[width=3.0in,trim=0 0 0 0,clip=true]{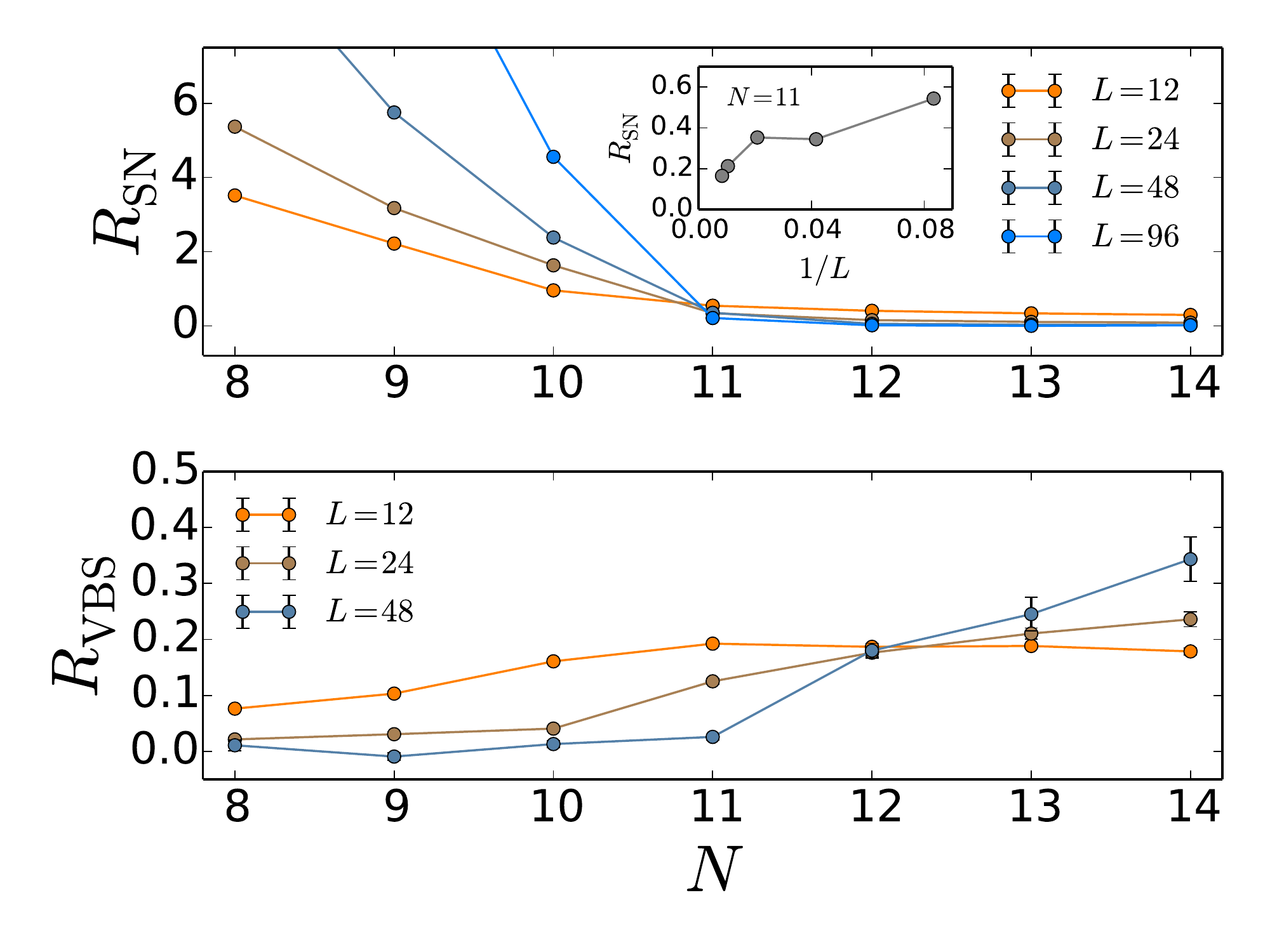}
\caption{\label{fig:R_MV} Crossing plots of the ratios $R_{\rm
    SN}$ and $R_{\rm VBS}$ as a function of the discrete variable
  $N$ for the $\hat H_J$ model. It is seen that spin nematic order is present for
$N\leq 10$. VBS order on the other hand is present for $N>
12$. $N=12$ appears to be on the verge of VBS order. Interestingly, $N=11$ has no SN
or VBS order. In the text, we present evidence that this phase is a
QSL. The inset in the upper panel shows $R_{\rm SN}$ scales to zero at $N=11$, despite non-monotonic behavior at intermediate $L$.}
\end{figure}

As shown in Fig.~\ref{fig:Sk}, a peak
in $S_{\rm SN} ({\bf
k})$ is found at the $\Gamma$ point. Comparing the data at $N=10$
and $N=14$, already qualitatively it is possible
to see the peak in $S_{\rm SN} ({\bf k})$ softens as $N$ is
increased.  In contrast  $\chi_{\rm VBS} ({\bf
k})$ develops sharp
peaks at the X and M points as $N$ is increased. These are precisely the
momenta at which previous numerical studies of the triangular lattice quantum dimer
model Eq.~(\ref{eq:qdm})
have observed Bragg peaks~\cite{ralko2006:qdm}, validating the
large-$N$ mapping to  Eq.~(\ref{eq:qdm}) made earlier. To detect at which $N$, the magnetic
order is destroyed and the VBS
order first sets in, we study the ratio, $R_{\rm SN} = 1 - \frac{S_{\rm
    SN}({\bf \Gamma + a}2\pi/L)}{S_{\rm SN}({\bf \Gamma})}$ (where
${\bf a}\equiv {\bf x}- {\bf y}/\sqrt{3}$) as a function of $L$. $R_{\rm SN}$
must diverge in a phase in which the Bragg peak height scales with
volume and becomes infinitely sharp. On the other hand it must go to zero in a phase in which
the correlation length is finite and the height and width of the
Bragg peak saturate with system size. At a critical point standard finite size scaling arguments
imply that the ratio, $R_{\rm SN}$ becomes volume independent. All of
these facts together imply a crossing
in this quantity for different $L$. Fig.~\ref{fig:R_MV} shows the
$R_{\rm SN}$ and $R_{\rm VBS}$ ratios (an analogous quantity constructed for the VBS
order from $\chi_{\rm VBS}({\bf k})$ close to the {\bf M}-point) as a function of the
discrete variable $N$ for different
$L$. The data for $R_{\rm SN}$ shows that the magnetic order
is present for $N\leq 10$. The $R_{\rm VBS}$ data shows that the 
long-range VBS order is present for $N> 12$. From Fig.~\ref{fig:R_MV}
we find that
 $N=12$ is on the verge of developing VBS order; from the system sizes
 accesible we are unable to
 reliably conclude whether $N=12$ has long range VBS order or not from our
 study. However, taken together the data show definitively that $N=11$ has neither VBS nor SN
order. As we shall substantiate below, at $N=11$, $\hat H_J$ is a quantum spin-liquid (QSL).

\begin{figure}[!t]
\includegraphics[width=3.5in,trim=10 80 0 100,clip=true]{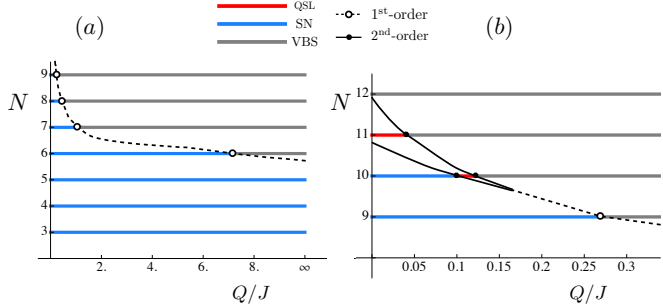}
\caption{\label{fig:pd} Phase diagram of $\hat H_{JQ}$   [Eqs.~(\ref{eq:projmodel},\ref{eq:qmodel})]
  for different values of $N$. The left panel shows the phase diagram
  for  small $N$, where a first order SN-VBS
  transition is found for $6\leq N \leq 9$, (see
  Fig.~\ref{fig:first_N7}). As $N$ is increased we find the first order transition
  weakens.  The right panel shows how an intermediate QSL phase emerges for $N=10$ and $N=11$. Transitions from the
  QSL to both SN and VBS phases are continuous on the large systems studied, see Fig.~\ref{fig:N10_cr}. }
\end{figure}

{\em J-Q models:} 
In order to
clarify the global phase diagram of SO($N$)
anti-ferromagnets and access the quantum phase transitions
between the SN, VBS and QSL phases found in $\hat H_J$, it is of interest
to find an interaction that can tune between these phases at {\em
  fixed} $N$. In order to be meaningful, the new coupling must preserve all the symmetries of
$\hat H_J$. To this end, we
introduce and study a 
generalization of the four-site $Q$ term of SU(2) spins~\cite{sandvik2007:deconf},
\begin{equation}
\label{eq:qmodel}
\hat H_Q = - Q \sum_{\langle ijkl \rangle }
\left ( \hat P_{ij} \hat P_{kl} + \hat P_{il}\hat P_{jk} \right )
\end{equation}
where the sum includes elementary plaquettes of length four on the triangular
lattice (with periodic boundary conditions on an $L\times L$ system
there are $3L^2$ such plaquettes).
For a fixed-$N$, $\hat H_Q$ provides
a tuning parameter which preserve both the internal and lattice symmetries of $\hat H_J$ and hence
allows us to study the generic phase diagram of SO($N$) magnets. A
summary of the phase diagram of $\hat H_{JQ}$ in the
$N$-$Q/J$ plane is in
Fig.~\ref{fig:pd}: The $Q$-interaction destroys
the SN order and gives way to VBS order only for
$N\geq 6$. We have found evidence for direct first-order SN-VBS transitions
for $6\leq N<10$ and exotic continuous SN-VBS transitions for $N=10$
and $N=11$.

\begin{figure}[!t]
\includegraphics[width=3.0in,trim=200 0 0 0,clip=true]{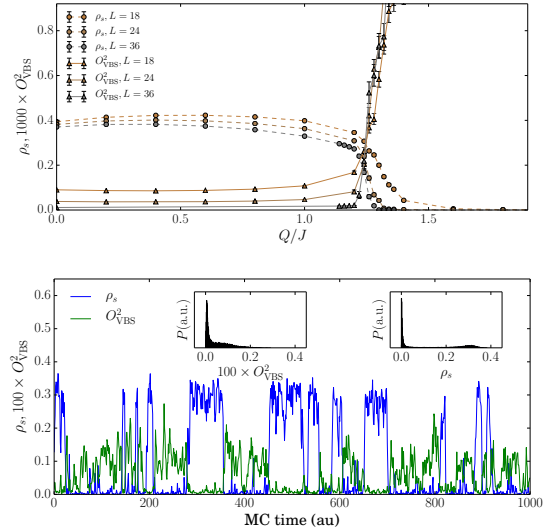}
\caption{\label{fig:first_N7} First-order SN to VBS 
  transition in $H_{JQ}$ at $N=7$. The upper panel shows the VBS order 
parameter and the stiffness as a function for $Q/J$ for different 
$L$ indicating a direct SN-VBS transition. The lower panel 
shows MC histories (and histograms in the inset) at $Q/J=1.26$,
providing clear evidence that the  SN-VBS transition at $N=7$
is direct and first-order. }
\end{figure}

As an example of our
observed first-order behavior we present in Fig.~\ref{fig:first_N7},
our study of the $N=7$ QMC data for the spin stiffness $\rho_s\equiv 
\langle W_x^2\rangle/L$ (where $W_x$
is the winding number of the spin world lines), which acts as a
sensitive order parameter for the SN phase, and the VBS order
parameter $O^2_{\rm VBS}\equiv \chi_{\rm VBS}({\bf M})/N_{\rm site}$. Clear evidence for a direct first order SN-VBS
transition at $N=7$ is found.

\begin{figure}[!t]
\includegraphics[width=3.5in,trim=0 0 0 0,clip=true]{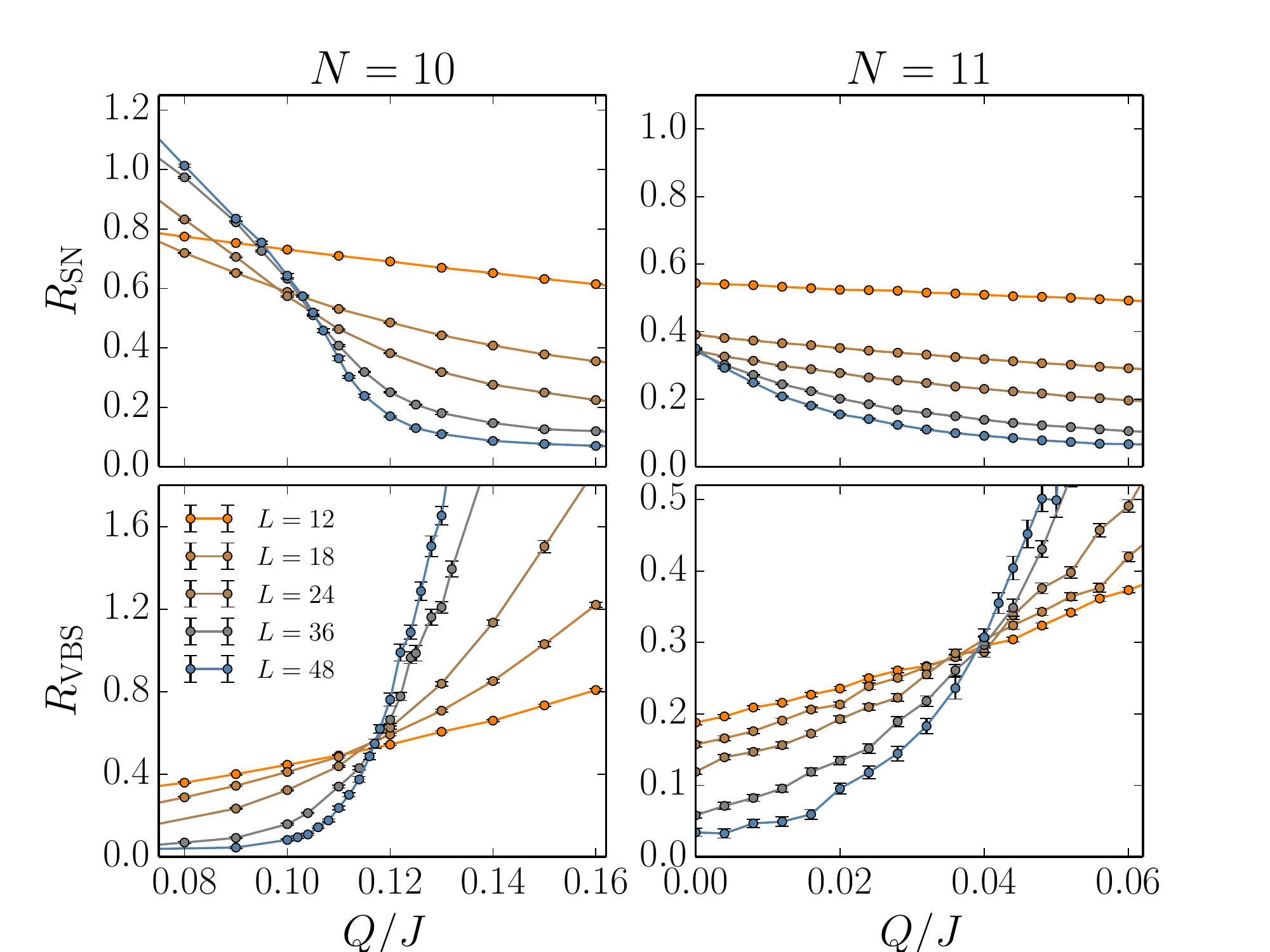}
\caption{\label{fig:N10_cr} Crossings of $R_{\rm SN}$ (above) and
  $R_{\rm VBS}$ (below) signaling the location of the onset of long-range SN
  and VBS orders at $N=10$ (left) and $N=11$ (right). At $N=10$, $R_{\rm
    SN}$ and $R_{\rm VBS}$ cross at close but significantly different couplings,
  $Q_c=0.100(5)$ and $Q_c=0.117(2)$ respectively. At $N=11$, $R_{\rm
    SN}$ appears to have crossed at $Q/J<0$ (we cannot study this
  region because of the sign problem), whereas $R_{\rm
    VBS}$ crosses at $Q_c=0.042(3)$. From the
  location of the crossings, for both $N=10$ and $N=11$, we can infer an intermediate phase which is
  neither SN nor VBS, as shown in Fig.~\ref{fig:pd}(b).  We present arguments that this phase
is a QSL. No direct evidence for
  first order behavior is found at either of the transitions, though a
  weakly first order SN-QSL cannot be ruled out. The QSL-VBS
  transitions shows good scaling behavior with unconventional critical exponents. }
\end{figure}

The nature of the transition changes at $N=10$, where evidence for two
phase transitions is found. As shown in Fig.~\ref{fig:N10_cr} the SN
order vanishes at a $Q/J$ smaller than the value at which VBS order
develops. Although the difference is small for $N=10$, it is significant.
 The data for $N=11$ in Fig.~\ref{fig:N10_cr} shows that the SN and VBS orders do not vanish at the same point. In fact $R_{\rm SN}$ indicates that
the SN order has vanished already at $Q/J=0$,
consistent with our previous analysis of $\hat H_J$. As illustrated by the
dashed and solid lines in Fig.~\ref{fig:pd}, the appearance of the QSL phase is
consistent with a global phase diagram for the SO($N$) magnets.

{\em QSL phase and criticality:} We have identified the 
ground state between SN
and VBS as a QSL, since it does not show evidence for any Landau-order. Were the intermediate phase
characterized by a conventional order parameter, we would have expected
strong first order transitions of the kind between
SN and VBS (see Fig.~\ref{fig:first_N7}), instead we find continuous transitions. 

There are field theoretic
reasons to expect a QSL on quantum disordering a
spin nematic. The long-distance description of
our SO($N$) models is given by a RP$^{N-1}$ theory (in contrast
to the CP$^{N-1}$ description of SU($N$) models~\cite{read1990:vbs}), which can be
described as $N$ real matter fields coupled to a $Z_2$ gauge field. Such a
theory is expected to host three phases~\cite{lammert1993:nematic}, a symmetry breaking phase in which
the matter condenses (which we identify in our spin model as the SN), a stable phase in which the matter gets a gap and
the $Z_2$ gauge theory is deconfined (identified here as the QSL) and
a phase in which matter is gapped and the $Z_2$ is confined
(identified here as the VBS). Thus, the SN-QSL critical point should
be in the universality class of O($N$)$^*$ critical point~\cite{xu2012:qcp}. The QSL-VBS phase transition should be in the same
universality class as the critical point between these identical phases in the
quantum dimer model since the magnetic fluctuations are gapped in both
the QSL and VBS phases. A previous analysis of this phase transition has predicted
an O(4)$^*$ phase transition~\cite{moessner2001:isfr}, where the VBS order parameter is
identified with a bilinear of the primary field. 

A detailed study of the critical phenomena at $N=10$ and $N=11$ is clearly beyond the scope
of the current manuscript. We shall be satisfied here with a brief analysis: At the QSL-VBS critical point, we are able to carry out reasonable data collapses~\cite{nbdrct2015:supmat} at both
$N=10$ and $N=11$ for $O^2_{\rm VBS}$
(for both X and M ordering vectors, see Fig.~\ref{fig:Sk}) and $R_{\rm
VBS}$, where we find, $\eta_{\rm VBS}=1.3(2)$ and $\nu_{\rm VBS}=0.65(20)$ for the anomalous
dimension of $O_{\rm VBS}$. The unusually large value
of $\eta_{\rm VBS}$ is a direct consequence of
fractionalization in the intermediate QSL phase and is often
regarded as a smoking gun diagnostic of exotic critical points (see
e.g., ~\cite{senthil2004:science}). More quantitatively, our critical exponents
are in rough agreement with the best estimate of
$\eta =1.375(5)$ of the bilinear field and $\nu=0.7525(10)$ in the O(4) model~\cite{Ballesteros1996:On}. We note that the
values for $\eta_{\rm VBS}$ and $\nu_{\rm VBS}$ agree within the quoted errors for $N=10$ and
$N=11$. Taken together, this bolsters the case that the intermediate
QSL phase has $Z_2$ fractionalization, albeit more work is needed for
a definitive identification. 
 Unfortunately, the SN-QSL transition,
observed only at $N=10$,
has large corrections to scaling and we are unable to reliably
determine its critical exponents or determine whether it is a weakly
first order transition (no direct evidence for a first-order
transition has been found of the type shown for the $N=7$ case).

In summary, we have introduced a new family of sign-free SO($N$) spin models, which can
be regarded as non-bipartite generalizations of their popular SU($N$)
cousins. The triangular lattice model which we have studied thoroughly
here hosts  a
spin nematic, a VBS with a large unit cell, a quantum spin liquid
phase and unusual quantum critical points. The absence in the SO($N$) models of a direct
continuous ``deconfined quantum critical point''~\cite{senthil2004:science} is in striking contrast
to previous simulations of the related bipartite SU($N$) models~\cite{kaul2013:qmc,block2013:fate}. We
have offered a plausible field theoretic scenario that naturally
explains this difference. It is interesting that the
absence (presence) of a QSL in bipartite SU($N$) (non-bipartite
SO($N$)) spin models seems to track the absence or presence of this phase in
the kind of quantum dimer models that our model maps to at large-$N$~\cite{moessner2001:rvb}.

While the study in this paper has focussed on the triangular
lattice,  our family of models, Eq.~(\ref{eq:projmodel},\ref{eq:qmodel}) may be
constructed sign free on any two or three dimensional non-bipartite lattice.  Because of the larger degree of frustration, the kagome system may provide a wider swath of the QSL
phase and hence could possibly allow a more detailed study of this
phase, even if the phase diagram is of the same form found here. Exploring the phase diagram and quantum
phase transitions of the three dimensional pyrochlore system
is an exciting open direction for future work. 

The author is grateful to J. Chalker, T. Lang, M. Levin, R. Mong,
G. Murthy, A. Nahum, A. Sandvik, T. Senthil and M. Zaletel
for helpful discussions. This research was supported in part by NSF DMR-1056536.


\putbib[/Users/rkk/LAPTOP/OPPIE/Physics/PAPERS/BIB/career.bib]

\end{bibunit}

\clearpage

\begin{bibunit}[apsrev]

\section{ SUPPLEMENTARY MATERIALS}

\subsection{Model and Symmetries}

Here we provide some additional details of the models introduced in
Eq.~(\ref{eq:jmodel}) and Eq.~(\ref{eq:projmodel}).

\subsubsection{Mapping between Eqs.~(\ref{eq:jmodel}) and~(\ref{eq:projmodel})}
 
To see the connection between the two Hamiltonians
Eq.~(\ref{eq:jmodel}) and Eq.~(\ref{eq:projmodel}).  We consider two
SO($N$) spins. We can combine them into three representations: a
singlet (S), symmetric ($\chi$) and anti-symmetric ($\Phi$)
representations of dimensions: 1, $\frac{N^2}{2} + \frac{N}{2}-1$ and
$\frac{N^2}{2}-\frac{N}{2}$. Now construct projectors on these
representations, $P_S,P_\chi$ and $P_\Phi$. Clearly
$P_S+P_\chi+P_\Phi=1$ and $P_S^2=P_S ,P_\chi^2=P_\chi ,P_\Phi^2=P_\Phi$. It is straightforward to show
that, $\left ( L_i \cdot L_j\right ) = -(N-1) P_S - P_\chi+P_\Phi$
by explicitly acting on the symmetrized wave-functions. From this it
follows that $\left ( L_i \cdot L_j\right )^2 = (N-1)^2 P_S +
P_\chi+P_\Phi$. From which it follows that $\left ( L_i \cdot
L_j\right )^2 = \left ((N-1)^2-1\right ) P_S + 1$, which proves as claimed that
for $N\geq 3$, Eq.~(\ref{eq:jmodel}) and Eq.~(\ref{eq:projmodel})
are equivalent up to a constant.

\subsubsection{$N=2$}

Although not studied in this manuscript, for the sake of completeness, we discuss
our model at $N=2$. Even though Eq.~(\ref{eq:jmodel}) is trivial for
$N=2$ (since
squaring the only SO(2) generator is just an identity operator),
Eq.~(\ref{eq:projmodel}) is a well defined non-trivial model.
Identifying the two colors with $\uparrow$ and $\downarrow$ spins, Eq.~(\ref{eq:projmodel}) becomes $\hat H  = \sum_{\langle ij \rangle}
-J( S^x_{i}S^x_{j}+S^z_{i}S^z_{j} )+ J S^y_{i}
S^y_{j}$. Previous work on the triangular lattice
$N=2$ model has
found clear evidence for SO(2) symmetry breaking superfluid order ~\cite{melko2005:tri,heidarian2005:tri,wessel2005:tri}.

\subsubsection{Symmetries}

We now discuss the symmetries of the model
Eq.~(\ref{eq:projmodel}). We begin by observing that this model is invariant under
uniform O($N$) rotations where we multiply each basis state by an
orthogonal matrix (one that satisfies $O^TO =1$), since this leaves the singlet state invariant, {\em
i.e.} 
\begin{equation}
\label{eq:ortho}
\sum_{\alpha}|\alpha \alpha \rangle \rightarrow \sum_{\alpha} O_{\alpha \gamma}O_{\alpha \eta} |\gamma \eta \rangle = \sum_{\alpha}|\alpha \alpha \rangle. 
\end{equation}
However we should identify rotations that only differ by changing all the local basis states by the same phase (in this
case a sign). Here it becomes necessary to distinguish between even
and odd $N$. This is because the matrix $-1$ has determinant 1 for even
$N$ and -1 for odd $N$. Thus for odd-$N$ the symmetry
is simply SO($N$), since the rest of O($N$) is obtained from SO($N$)
by multiplying by -1. For even-$N$ however SO($N$) has pairs of elements
that cause the same basis transformation up to a sign, {\em e.g.} 1
and -1. On the other
hand unlike the case of odd-$N$, the O($N$) matrices with determinant -1 are independant symmetries,
so the symmetry realized for even-$N$ is an $\frac{O(N)}{Z_2}$.

\begin{table}[t]
\begin{tabular}{||c||c|c|c|c|c|c||} 
\hline 
\hline
size & $N$ & $Q$  & $\beta_{\rm QMC}$ & $E_{ex}$ & $E_{\rm QMC}$  \\
\hline
\hline $2\times 2$ & $4$ & $0$ &16& $-1.5$ & $-1.49997(3)$\\
\hline $2\times 2$ & $4$ & $1$ &16& $-4.5$ & $-4.5000(1)$\\
\hline $2\times 2$ & $5$ & $0$ &16& $-1.4$ & $-1.40000(5)$\\
\hline $2\times 2$ & $5$ & $1$ &16& $-4.2$ & $-4.2001(1)$\\
\hline $2\times 3$ & $3$ & $2$ &16& $-6.0657499233$ & $-6.0656(1)$\\
\hline $3\times 3$ & $2$ & $0$ &16& $-1.7026987262$ & $-1.70269(2)$\\
\hline $3\times 3$ & $2$ & $1$ &16& $-3.7290340614$ & $-3.72900(3)$\\
\hline $3\times 3$ & $2$ & $2$ &16& $-5.7639923092$ & $-5.76402(5)$\\

\hline

\hline
\hline 
\end{tabular}
\caption{Test comparisons of ground state energies from exact
  diagonalization and average energies from
 finite-$T$ QMC studies of the SO($N$) model introduced here.  Note that $J=1$ always. The energies reported here are per site and
on triangular lattices with periodic boundary conditions such that
there are always $3L^2$ bonds and $3L^2$ plaquettes in
Eqs.~(\ref{eq:projmodel},\ref{eq:qmodel}). This causes some terms to
be appear more than once for the $2\times 2$ and $2\times 3$ systems.}
\label{tab:qN}
\end{table}

\subsubsection{Symmetry on Bipartite Lattices}

On bipartite lattices the orthogonal rotation symmetry,
Eq.~(\ref{eq:ortho}) gets extended to a unitary symmetry (with
$U^\dagger U =1$) so long as the
singlet is defined between sites on opposite sub-lattices, and A
sub-lattice spins are rotated by $U$ and B sub-lattice spins are
rotated by $U^*$,
\begin{equation}
\label{eq:unitary}
\sum_{\alpha}|\alpha \alpha \rangle \rightarrow \sum_{\alpha} U^*_{\alpha \gamma}U_{\alpha \eta} |\gamma \eta \rangle = \sum_{\alpha}|\alpha \alpha \rangle. 
\end{equation}
Since a uniform phase change of the all the states locally does not
have physical consequences, the model is said to have an
SU($N$) symmetry, as has been discussed and extensively studied
previously in such models, see {\em e.g.} Ref.~\cite{kaul2013:qmc} for a review.

\subsubsection{Ground state theorems}

Marshall's sign
theorem guarantees that the ground state of $H_{JQ}=H_{J}+H_{Q}$ is an SO($N$) singlet. In addition, on the
triangular lattice, which is the focus of our study here,  there is no
simple translationally invariant covering of two-site singlets, leading us to
suspect that a generalization of the SU(2) square lattice Lieb-Schultz
Mattis (LSM) theorem~\cite{hastings2004:lsm}
applies to $H_{JQ}$ on this lattice, {\em i.e.} in the thermodynamic limit there must
be a degeneracy in the ground state, so that a simple gapped paramagnet is not possible --
 either a symmetry is broken or the ground state is exotic. A rigorous
 proof of this
intuitive assertion is expected to be at least as technical as the
proof for the bipartite $N=2$ case~\cite{hastings2004:lsm} and is
beyond the scope of this work. As we saw above, on a one-dimensional
chain, which is bipartite,
our model is equivalent to the SU($N$) model studied by
Affleck~\cite{affleck1985:lgN} and is hence expected to have an LSM degeneracy.

\subsection{Numerical Simulations}

\subsubsection{ QMC energy tests}

Here we provide the results of some QMC tests on small lattices for the total energy
per spin of our models, $\hat H_J+H_Q$, Eqs.~(\ref{eq:projmodel},\ref{eq:qmodel}) on the triangular lattice, for completeness and future
comparisons.

\begin{figure}[!t]
\includegraphics[width=3.5in,trim=0 0 0 0,clip=true]{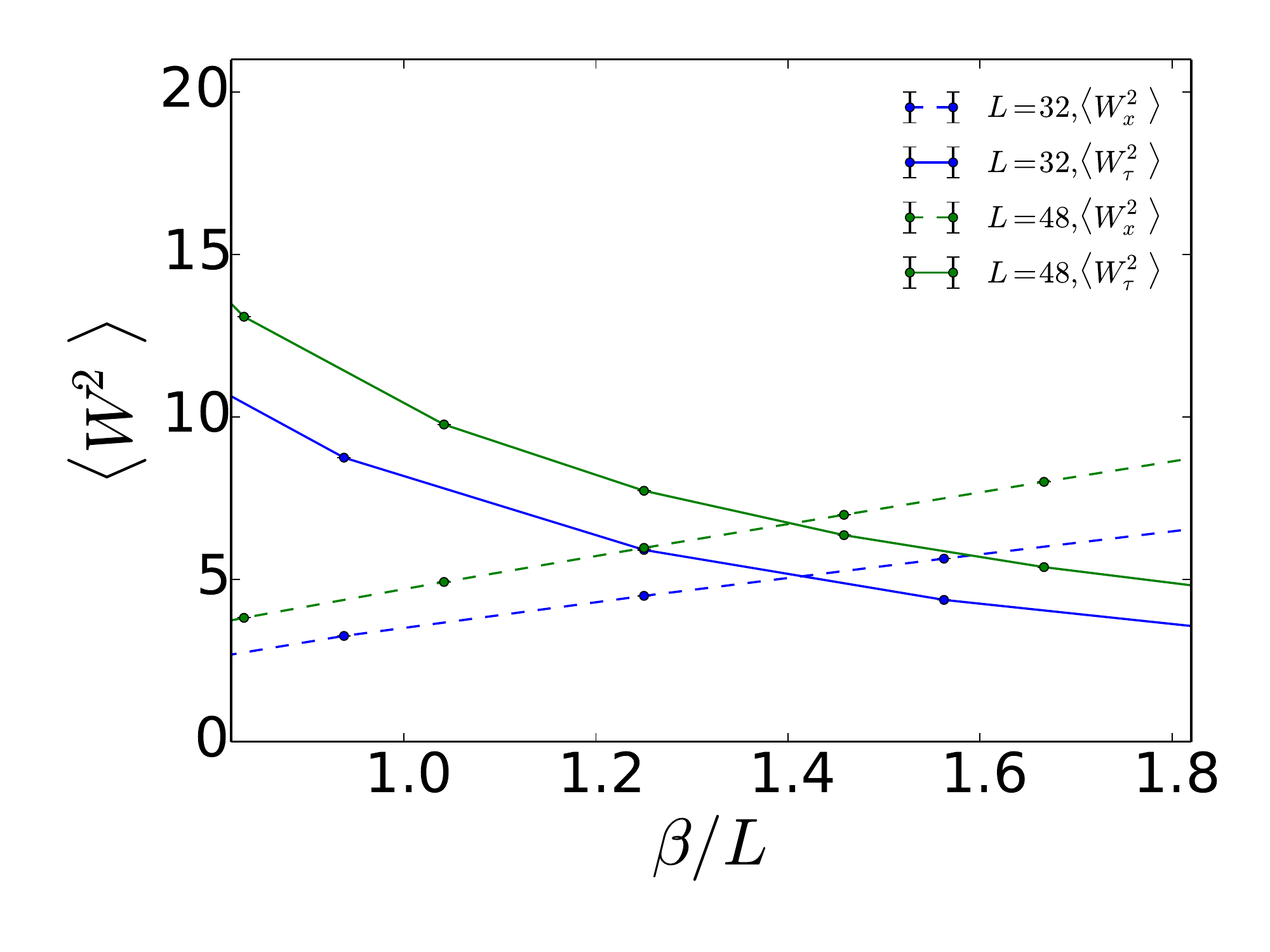}
\caption{\label{fig:BL} Estimation of optimal $\beta/L$ ratio for
  our simulations by comparison of the fluctuations of the temporal
$(\langle W_\tau^2\rangle)$ and spatial   $(\langle
W_x^2\rangle)$ winding numbers
. Data shown is for $\hat H_J$ at $N=10$.  }
\end{figure}

\subsubsection{Choice of $\beta=L$}

In Fig.~\ref{fig:BL} we show the dependence of the fluctuations of the temporal and
spatial winding numbers on the ``aspect ratio'' of our simulation
cell. $\beta$ is a measure of the extent of the imaginary time and $L$
is an estimate for the linear spatial extent. We study the
fluctuations of the temporal and spatial winding numbers as the ratio
$\beta/L$ is varied for two different sizes, $L=32$ and $L=48$ at
$N=10$ in the model $\hat H_J$. We
find that both quantities are balanced at a value of $\beta/L$ which
is of the  order of one (close to 1.42) and that the crossing point
does not move much with system size. Thus for simplicity we have chosen
$\beta/L=1$ throughout the paper.

\begin{figure}[!h]
\includegraphics[width=3.5in,trim=0 0 70 100,clip=true]{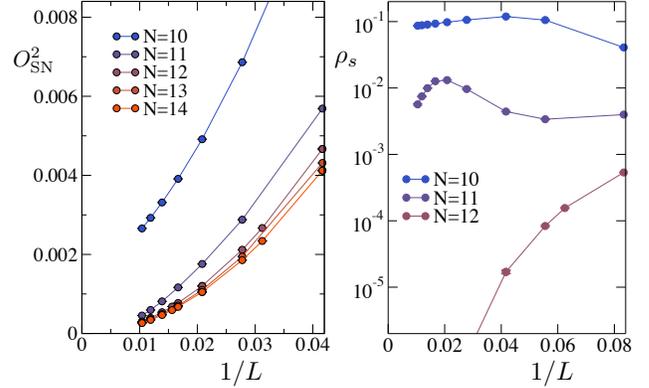}
\caption{\label{fig:mag_peaks} Finite size scaling of spin nematic order 
  parameter and spin stiffness for the model $\hat H_J$, in Eq.~(\ref{eq:jmodel}) with 
  $10\leq N \leq 14$. Shown on the left is the 
  square of the spin nematic order parameter, $O_{\rm SN}^2$ (the height of the 
  Bragg peak in Fig.~\ref{fig:Sk}), and on the right is the spin 
  stiffness, $\rho_s$, plotted as a function of $1/L$. The data 
confirms that the system is magnetically ordered for $N\leq 
  10$ and non-magnetic for $N\geq 11$.}
\end{figure}

\begin{figure}[!h]
\includegraphics[width=3.5in,trim=0 0 0 0,clip=true]{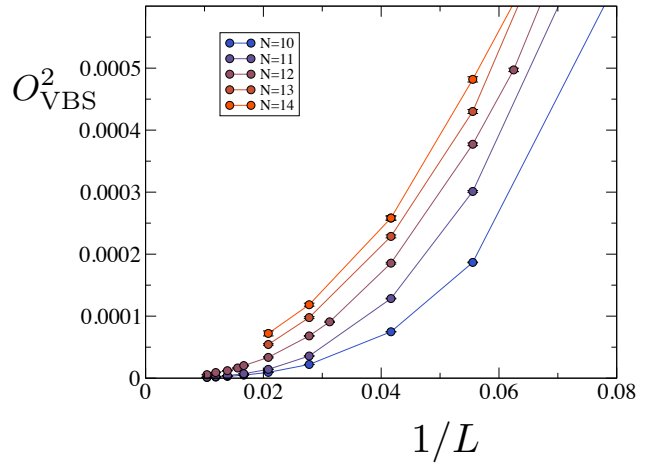}
\caption{\label{fig:vbs_peaks} Finite size scaling of VBS order
  parameter for the model $\hat H_J$, in Eq.~(\ref{eq:jmodel}) with 
  $10\leq N \leq 14$. For $N\geq 13$ we encounter difficulties in equilibrating the
  system for sizes larger than $L=48$ due to formation of long-range
  VBS order. Rather than extrapolating $O^2_{\rm VBS}$, we study
  the crossing of the ratio $R_{\rm VBS}$, which
  provides a reliable way to detect the onset of long-range VBS order
  on moderate system sizes [see Fig.~\ref{fig:R_MV}].}
\end{figure}

\subsubsection{Extrapolation of order parameters}

The simplest estimate for long range order is to study whether the
height of the Bragg peak per unit volume extrapolates to a finite
quantity in the thermodynamic limit. Unfortunately, this method becomes increasingly
unreliable when the measured order is weak, e.g., close to a critical
point. In such cases, results from extrapolations will depend on the form of the
extrapolation used. A thorough discussion of these difficulties in quantum spin
systems may be found in the literature~\cite{sandvik2012:boundary}. It
is for this reason that we prefer to work with the $R$ ratios defined
in the text. The disadvantage is that we do not know the order
paramater in the thermodynamic limit, but the advantage is we can calculate the
critical coupling reliably by studying the crossing of the $R$ ratio. For completeness we present here the data required for
extrapolation of both SN and VBS order parameters for $\hat H_J$. To test quantitatively for long range order we study the scaling of
the height of
the peak in $S_{\rm SN} ({\bf k})$, $O_{\rm SN}^2\equiv S_{\rm M} ({\bf
  k}=0)/N_{\rm site}
$ and the spin stiffness $\rho_s$ on finite size systems with $N_{\rm
  site}=L\times L$. Both quantities are expected to be finite in the
M state and zero when the O($N$) symmetry is
restored. Fig.~\ref{fig:mag_peaks} shows finite size data for both
quantities for different values of $N$. From these plots we conclude
that the M symmetry is broken up to $N=10$ and is restored for $N\geq 11$,
because $O_{\rm SN}^2$ scales to zero for these $N$. This behavior is
mirrored in $\rho_s$, albeit for intermediate $L$ there is some
non-monotonic behavior for $N=11$. This is consistent with our
conclusions in the main text made from the analysis of $R_{\rm SN}$

Finite size scaling for  the VBS order parameter is shown in
Fig.~\ref{fig:vbs_peaks}. Notice for the cases where there is VBS
order ($N=13,14$) we only have data for $L\leq 48$. For system sizes
larger than this we face serious equilibration issues with QMC as is
expected, since the simulation gets locked into a symmetry broken VBS
state. The plot serves to illustrate te ambiguity faced by making
direct extrapolations. On the other hand, a study of the $R$ ratios
shown in Fig.~\ref{fig:R_MV} provides a more clear cut way to locate
the critical point.

\begin{figure}[!t]
\includegraphics[width=3.5in,trim=0 0 0 0,clip=true]{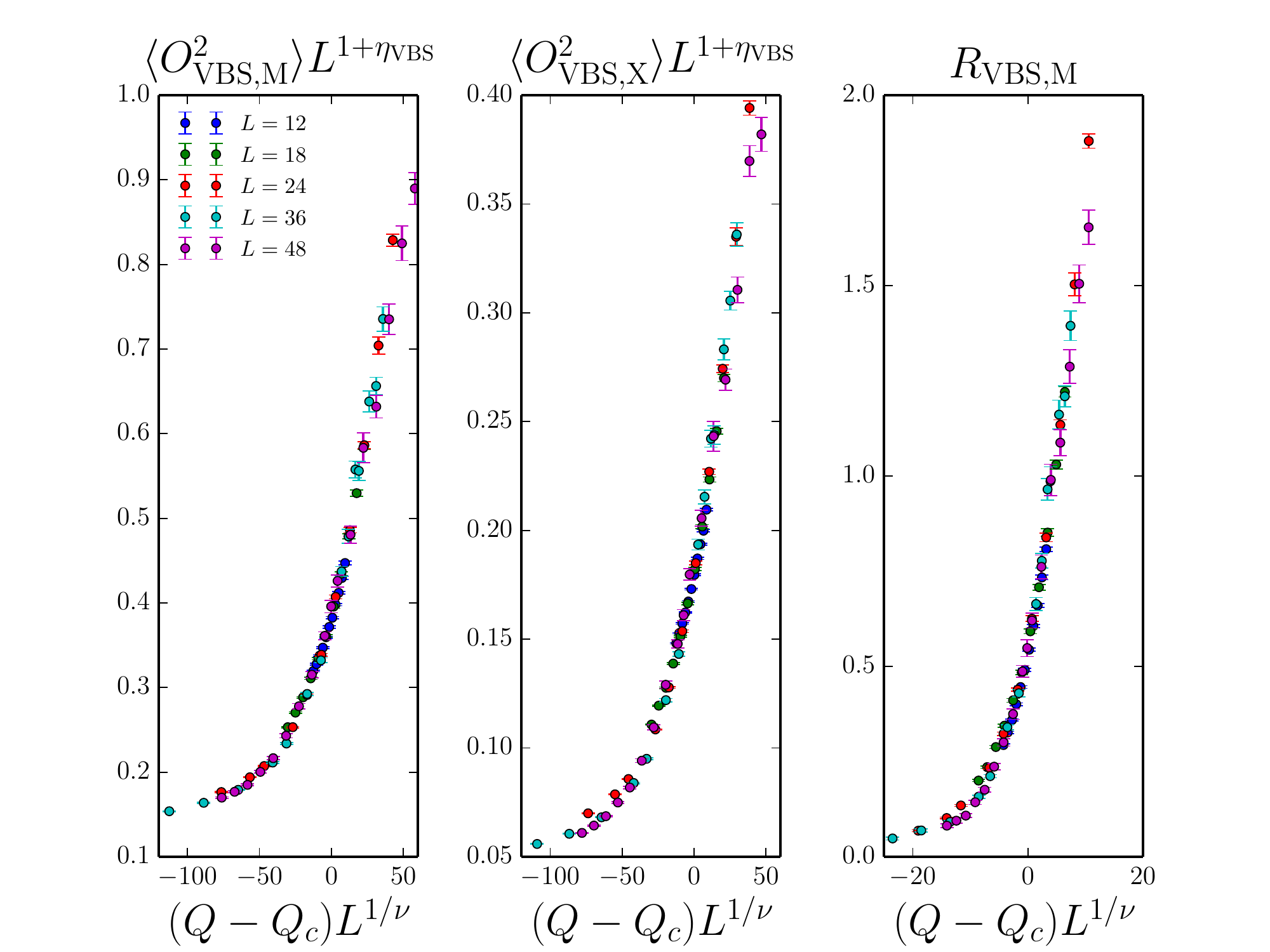}
\caption{\label{fig:clps10} Examples of our data collapses at QSL-VBS critical point at
  $N=10$ shown here with the best parameters determined individually
  for each observable. The parameters used are $Q_c=0.1170,0.1187,0.1171$, $1/\nu_{\rm
    VBS}=2.171,2.152,1.733$ and  $\eta_{\rm
    VBS}=1.414,1.106,-$. }
\end{figure}

\begin{figure}[!h]
\includegraphics[width=3.5in,trim=0 0 0 0,clip=true]{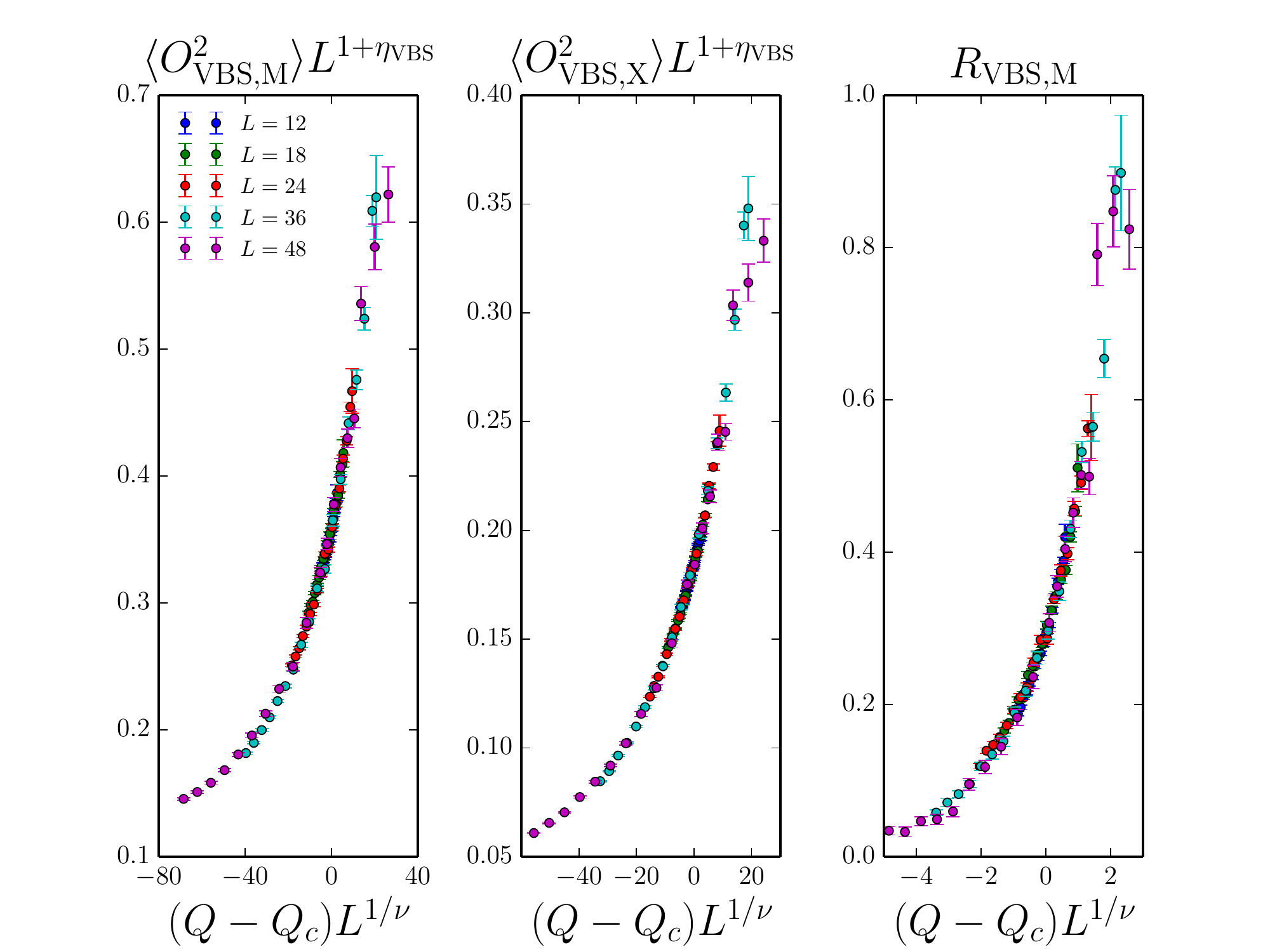}
\caption{\label{fig:clps11} Examples of our data collapses at QSL-VBS critical point at
  $N=11$ shown here with the best parameters determined individually
  for each observable. The parameters used are $Q_c=0.04330,0.04182,0.03917$, $1/\nu_{\rm
    VBS}=1.902,1.858,1.244$ and  $\eta_{\rm
    VBS}=1.383,1.133,-$. }
\end{figure}

\subsubsection{QSL-VBS and SN-QSL phase transitions}

Here we present some details of the study of the both the QSL-VBS ($N=10,11$) and
SN-QSL ($N=10$) phase transitions found in our model.

We obtain critical exponents at the QSL-VBS critical point by
attempting a data collapse, see Fig.~\ref{fig:clps10},\ref{fig:clps11}. We use  the standard
finite size scaling ansatz for the order parameter and the crossing ratio,
\begin{eqnarray}
\langle O_{\rm VBS}^2 \rangle &=& L^{-(1+\eta_{\rm VBS})}{\mathcal F}_O (g 
  L^{1/\nu_{\rm VBS}})\\
R_{\rm VBS} &=& {\mathcal F}_R (g 
  L^{1/\nu_{\rm VBS}})
\end{eqnarray}
where $g=(Q-Q_c)/J$. We continue to work with $\beta=L$ as
discussed. No attempt is made to make use of corrections to this
leading scaling behavior. Our main objective is to determine the
universal number $\eta_{\rm VBS}$ for the QSL-VBS transition for $N=10$
and $N=11$. We find acceptable collapses for our data sets
over a wide range of
$\nu_{\rm VBS}$. On the other hand, the
estimate for $\eta_{\rm VBS}$ is relatively stable over our various
fits. The values and errors of the critical exponents quoted in the
main text are based on the variation observed by using different data
sets. A higher precision study should be possible with access to
more accurate data and larger system sizes. In order to carry out the collapse numerically, we make
use of a recently developed Bayesian approach to
scaling~\cite{harada2011:bay}. We note that difficulties in obtaining
accurate values of the critical exponents at exotic transitions in
quantum spin models is a well-documented difficulty~\cite{harada2013:deconf}.

\begin{figure}[!t]
\includegraphics[width=3.5in,trim=0 0 0 0,clip=true]{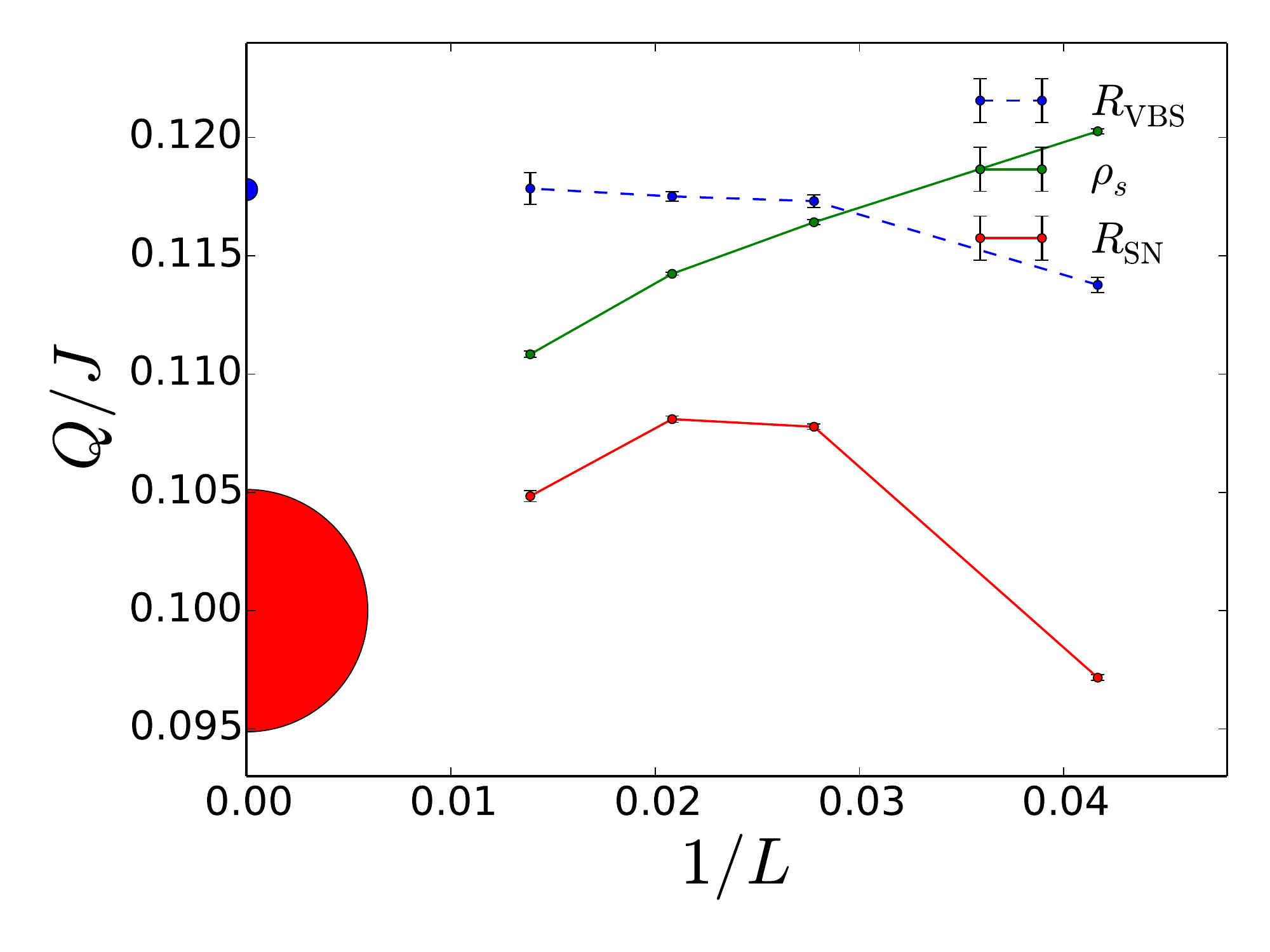}
\caption{\label{fig:O10}  Crossing of $L$ and $L/2$ for various dimensionless quantities for $N=10$. The dashed line is a quantity ($R_{\rm VBS}$) whose crossing locates the VBS transitions and the solid lines are for two independent quantities ($\rho_s$ and $R_{\rm SN}$) that locate the SN transition. The blue and red semi-circle shows the range of critical couplings based on the extrapolation of this data. Note that (1) there is clear evidence for an intermediate phase. (2) the errors in the SN transition are significantly larger than those for the VBS transition.}
\end{figure}

Another quantum phase transition takes place between SN and QSL. In
our model this transition appears only at $N=10$. In
Fig.~\ref{fig:O10} we study the drift of various crossing quantities
at the critical points.  Presumably the significant drift for the
crossing at the  SN-QSL transition are due to corrections to
scaling. We have looked for signs of first order behavior as we found
for smaller-$N$ and not found them here, though the possibility of a
very weak first order transition cannot be ruled out. The corrections
to scaling hamper efforts to extract critical exponents at this phase
transition. In contrast the QSL-VBS transition shows a reasonably
converged crossing point with a nice scaling regime, where the
crossing points do not depend significantly on $L$.

\putbib[/Users/rkk/LAPTOP/OPPIE/Physics/PAPERS/BIB/career.bib]

\end{bibunit}

\end{document}